\documentclass[a4paper]{jpconf}

\usepackage{graphicx}
\usepackage{citesort}
\usepackage{epsfig}
\usepackage{subfigure}

\begin{document}
  \title{Direct photons in Au+Au collisions measured with the PHENIX detector at RHIC}
  
  \author{Richard Petti}
  
  \address{Department of Physics and Astronomy, Stony Brook University, SUNY, Stony Brook, New York 11794, USA}
  
  \ead{richard.petti@gmail.com}
  
  \begin{abstract}
    A major goal of experiments in heavy-ion physics is the characterization of the quark gluon plasma (QGP) produced in the collision of heavy ions at high energy.  Direct photons are a particularly good probe of the produced medium because they do not interact strongly and so can escape the medium unmodified, carrying information about when the photon was produced.  It is expected that direct photon contributions from different sources (QGP radiation, hard scattering, hadron gas radiation) dominate at different transverse momentum ranges.  Low momentum direct photons are dominated by thermal radiation (both from the QGP and hadron gas), while high momentum direct photons dominantly come from hard parton scatterings in the initial collision.  We present a summary of techniques to measure direct photons with the PHENIX detector, with a focus on low momentum direct photons through their external conversion to dilepton pairs.
  \end{abstract}
  
  \section{Introduction}
  Experiments at the Relativistic Heavy Ion Collider (RHIC) at BNL have established evidence that a quark-gluon plasma has been produced in heavy-ion  collisions at $\sqrt{s_{NN}} = 200$ GeV~\cite{whitepaper}.  One signature amongst many is that the system appears to behave like a liquid with partonic degrees of freedom, as evidenced by elliptic flow measurements and number of constituent quark scaling~\cite{ellipticFlow}.  Another observation was the suppression of high momentum probes, indicating that the medium is not completely transparant to colored objects~\cite{jetSupp}.  Recent PHENIX measurements of the temperature of the plamsa also indicate a temperature well above the expected critical temperature needed to form a QGP predicted by lattice QCD calculations~\cite{ppg086}.  A major goal of the RHIC program is to now study and characterize this extreme state of matter.
  
  In this paper we focus on photons as probes of the QGP.  Photons do not interact strongly with the medium and so escape virtually unmodfied, carrying information about each stage of the collisions where the photon was created.  We are interested in direct photons, which are defined as all photons that do not originate from hadronic decays.  This includes thermal and prompt photons, along with bremsstrahlung and jet-conversion photons.  We mention two important pieces of information (although there are many others) that can be extracted from the study of direct photons. One is the temperature of the plasma, which can be accessed by the shape of the photon distribution.  Another is the elliptic flow of direct photons.  Theory calculations~\cite{thermal_v2_theory} indicate that the elliptic flow of thermal photons is quite sensitive to the thermalization time, $\tau_0$, of the plasma.  It is expected that a small thermalization time leads to a small $v_2$, as the proportion of photons coming from the QGP phase is larger for earlier thermalization times.  The shape of the $v_2$ as a function of $p_T$ also changes drastically depending on the assumed $\tau_0$.  Thermal photons are expected to dominate at low momentum, and so we focus on measuring direct photons below $5$ GeV.
  
  The measurement of low momentum direct photons is notoriously difficult due to very large background from hadron (mostly $\pi^0$) decays.  This fact, combined with the limited energy resolution of the electromagnetic calorimeter (EMCal) at low energy, requires an alternate method to simply measuring photons that directly deposit their energy in the EMCal.  Thus it is advantageous to measure dilepton pairs originating from the conversion of photons.  The momentum resolution in the tracking systems for charged particles improves as $p_T$ decreases.  We consider two sources of dilepton pairs from photons, internal and external conversions.  Internal conversions occur from processes that happen to produce a virtual photon, rather than a real photon, which will decay into a low mass $e^+e^-$ pair~\cite{ppg086}.  External conversions occur when a photon interacts with material in the experimental apperature and converts into the dilepton pair.  The external conversion analysis is complementary to the virtual (internal) conversion analysis and is a good cross-check of the validy of the procedure and is one of the main motivations for undertaking the external conversion analysis.
  
  \section{Experimental Apparatus}
  A detailed description of the PHENIX detector subsystems can be found in the reference~\cite{detectorOverview}.  Photons are measured in the central arms of the PHENIX spectrometer, each of which cover $\pi/2$ in azimuth and total cover a psuedorapidity range of $\eta < |0.35|$.  Charged particles are tracked in layers of multi-wire proportional chambers, calculating the bend in the magnetic field and hence the momentum of the track~\cite{centDet}.  A combination of the Ring-Imaging Cherenkov Dector (RICH)~\cite{pid} and the electromagnetic calorimeter (EMCal)~\cite{emcal} assist in electron identification.  There are forward and backward beam-beam counters (BBCs) at $3.1 < |\eta| < 3.9$, utilized as global detectors to determine both centrality and the vertex of the collision along the beamline (the z direction).  A dedicated reaction plane detector consisting of plastic scintillator paddles was also installed for the 2007 run~\cite{rxnpDet}, which reside at $1.0 < |\eta| < 2.8$.
  
  \section{Analysis Techniques}
  The details of the internal conversion analysis can be found in the reference~\cite{ppg086}.  The basic idea of this technique is that low mass $e^+e^-$ pairs are formed when a virtual direct photon is produced in the system.  Many of these pairings exist from pure combinatorics and have no physical correlation.  These are removed by studying pairs formed in mixed events.  Further sources of correlated background pairs must also be removed (such as external conversions, cross pairs, jet pairs).  These can be removed by studying like-sign foreground distributions.  Once all background is removed, the yield of pairs is compared to a cocktail of expected hadronic sources.  An excess of pairs is observed in Au+Au collisions and is interpreted as originating from virtual photons.  The yield of real photons can be calculated from the pair rate observed.  Upon calculating the cross-section of direct photons, an excess above the expected rate is seen below $p_T < 3$ GeV.  This excess is exponential in shape and is fit to extract a temperature parameter.  The temperature measured in the paper for central Au+Au collisions is $T = 221 \pm19(stat)\pm19(sys)$ MeV, which is above the expected critical temperature needed for the phase transition.  This is an important result and it is worthwhile to seek independent measurements of the same quantity as a cross check.  The external conversion channel provides us with this cross check.
  
  The details of the external conversion analysis are described here.  We start by identifying single electrons, which will be paired later.  The electron id detectors used are the RICH and EMCal using common PHENIX electron id cuts.  It is required that the track is associated with $> 2$ phototubes in the RICH.  It is also required that the energy to momentum ratio, $E/p > 0.5$.  The reconstructed momentum of the track must be $0.2 < p_T < 20$ GeV.
  
  We need to identify pairs that actually come from an external photon conversion.  During the 2007 RHIC run, the Hadron Blind Detector (HBD)~\cite{HBD_ref} was installed for its engineering run.  The backplane of the detector is fairly thick (about $4\% X_0$) and so is a significant localized source of conversions.  Hence we choose to focus on conversions occurring in this backplane. It is located at a radius of about $60$ cm from the interaction point.  This significant conversion radius is the key to indentifying these photon conversions.
  
\begin{figure}[h]
    \begin{minipage}{18pc}
      \includegraphics[scale = 0.37]{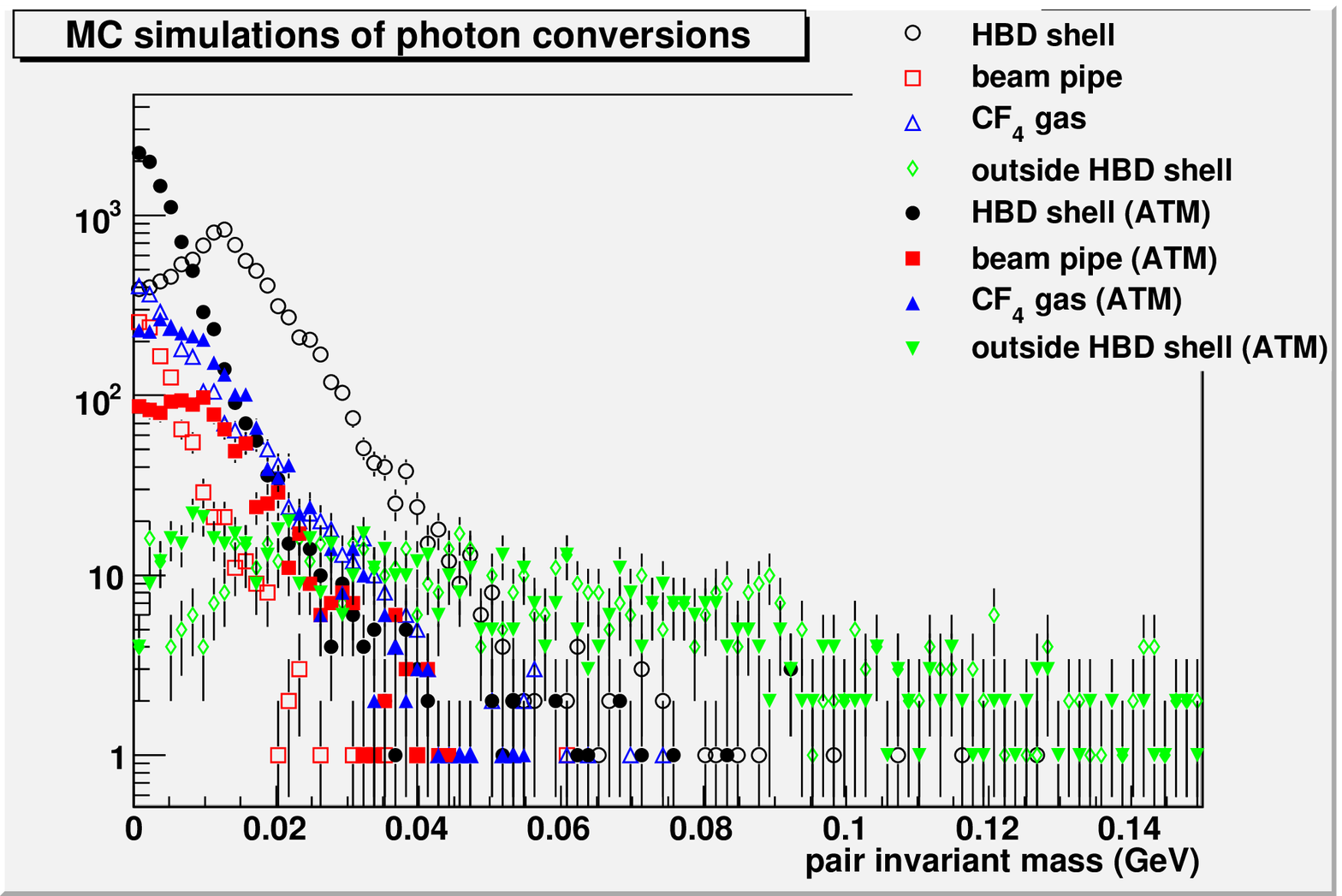}
      \caption{\label{fig:conversion_mass}The $e^+e^-$ pair invariant mass distriubtion for photon conversions at different radii is plotted from GEANT based Monte Carlo simulations.  The source of the conversion is represented by various shapes as shown in the legend.  Open symbols represent the normally reconstructed mass, with the closed symbols as the ATM calculated mass.}
    \end{minipage}\hspace{2pc}%
    \begin{minipage}{18pc}
      \includegraphics[scale = 0.38]{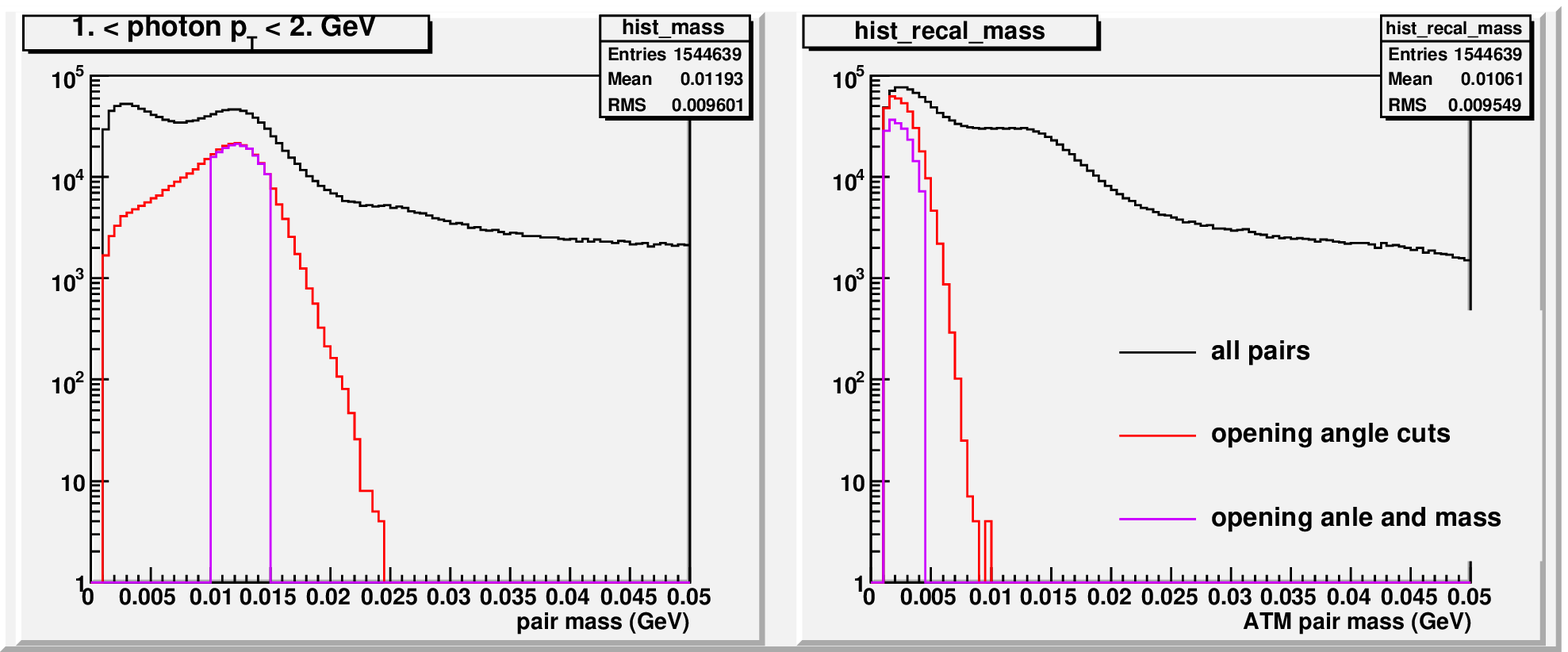}
      \caption{\label{fig:mass_with_cuts}The application of the pair cuts on the raw pair invariant mass distribution shown for one particular $p_T$ bin in real data.  The left side shows the normal track model and the right side the ATM.  Each color represents the application of more cuts as shown in the legend.}
    \end{minipage} 
  \end{figure}

  PHENIX does all of its tracking outside the magnetic field.  This requires an assumption about the origin of the track to determine the bend in the field and calculate the momentum of the particle.  A natural assumption is that tracks originate from the event vertex.  But if we consider electrons coming from a radius of $60$ cm, then this assumption is wrong and leads to a misreconstruction of the electron track.  This misreconstruction leads to an artificial opening angle of the pair, which leads to an apparent mass of the pair.  This apparent mass is proportional to the radius at which the conversion occurs (the controlling parameter is how much field the particle missed, even though it was assumed to pass through the entire field), causing a slight over-estimate of the momentum.  The apparent mass can therefore be used to help identify conversions at the HBD backplane.  Simualtion studies have shown that we expect an invariant mass peak of the dilepton pairs to be at $12$ MeV with the Run 7 magnetic field configuration, see the open symbols in Fig. \ref{fig:conversion_mass}.  This same peak can be seen in the data, Fig. \ref{fig:mass_with_cuts}.
  
  Of course it is desirable to have the correct momentum of the converted photon, not a misreconstructed momentum.  Therefore we correct for this mis-measurement with an alternate track model (ATM) assumption.  Rather than assuming that tracks originate from the event vertex, it is assumed that all the tracks originate from a radius of $60$ cm.  The momentum of the reconstructed track (under the normal track model assumption) is reparameterized for the ATM assumption.  This is studied through Monte Carlo simulations.  The conversion pairs are then reconstructed properly under this ATM assumption, and the invariant mass moves from peaking at $12$ MeV towards having a peak at $0$ MeV.  All other particles originating from the event vertex will now be misreconstructed, but this is a desired effect.  Pairs from particles originating from the event vertex will be shifted up in mass.  This is especially important for Dalitz decays of $\pi^0$'s, since the HBD conversion peak sits on the side of the Dalitz peak (see the left hand side of Fig. \ref{fig:mass_with_cuts}, where the first peak seen in the black historam is the Dalitz peak and the second peak from conversions at the HBD).  We cut on how the mass moves under the two track models, allowing us to indentify converted photons.  It is required that the ATM calculated mass is less than $4.5$ MeV and the normally calculated mass is between $10 < mass < 15$ MeV.
  
  In addition to these mass cuts, we also cut on the separation of the tracks at the HBD backplane calculated with the corrected ATM momenta.  Conversion pairs should have no opening angle at the point of conversion.  Therefore we require that the tracks are close in $\phi$ and $z$.  These cuts allow us to get a very clean photon id at low momentum.  The effect of these pair cuts on the raw $e^+e^-$ pair invariant mass distribution is shown in Fig. \ref{fig:mass_with_cuts}.  An unfortunate detector problem offers a good opportunity for an internal cross-check of the analysis, namely the removal of half of the HBD for repair during the run.  And so we can check the misidentification rate of the photons by analyzing the pairs we id in the arms serparately and comparing the number of photons that we find to convert in the arms.  It is observed that the misidentification of converted photons is less than $3\%$.  The combinatorial background is insignificant in this very low invarant mass range (mass less than $20$ MeV).
  
  The final goal of the measurement is the direct photon $v_2$ (at the time of writing this analysis is still a work in progress and so only an analysis strategy and current status is presented).  We need three basic pieces of information to derive the direct photon $v_2$.  We measure the $v_2$ of all photons (we cannot a priori distinguish the origin of a photon) and statistically subtract background sources to the $v_2$.  Therefore, we measure the inclusive photon $v_2$.  In addition to the inclusive photon $v_2$, we need to also know the $\pi^0$ decay photon $v_2$ (since most of the inclusive photons come from $\pi^0$ decays).  This is the only hadronic $v_2$ source that we measure.  Contributions to the $v_2$ from other (less significant) sources of hadrons will be calculated from $KE_T$ scaling, where $KE_T = m_T - m_0$.  We also need to know the relative fraction of direct photons compared to hadronic sources or $R_\gamma = N^{incl}/N^{hadron}$, the ratio of inclusive photons to hadronic decay photons.  Then the direct photon $v_2$ can be calculated from Eqn. \ref{eqn:directV2}.

  \begin{equation}
    v^{dir}_2 = \frac{N^{incl}v^{inc}_2 - N^{BG}v^{BG}_2}{N^{inc} - N^{BG}}  = \frac{R_\gamma v^{incl}_2 - v^{BG}_2}{R_\gamma - 1}
    \label{eqn:directV2}
  \end{equation}

  The inclusive photon $v_2$ is measured by taking the identified external photon conversions using the aforementioned method and correlating the angle of emission, $\phi$, relative to the reaction plane angle, $\Psi$, in azimuth.  This $\phi - \Psi$ distribution is binned in $p_T$.  Correlation functions (C.F.) are constructed from these distributions.  We perform a Fourier decomposition of the correlation functions, following Eqn. \ref{eqn:v2}.  The correlation functions are fit with Eqn. \ref{eqn:v2} and the elliptic flow strength, $v_2$, is extracted in each $p_T$ bin.  Higher order Fourier terms in Eqn. \ref{eqn:v2} are assumed to be negligible and are not included.  Sine terms are zero by symmetry.  We also calculate $v_2$ as $\left<cos(2(\phi-\Psi))\right>$, the mean projection method.  This yields consistent results with the fit and the average of the two is taken for the result presented here.  The extracted $v_2$ is corrected by the reaction plane resolution, determined by comparing correlations between the north and south detectors.  The major source of systematic error on the points is from the reaction plane determination, which is roughly $10\%$.

  \begin{equation}
    C.F. = B[1+2v_2cos(2(\phi-\Psi))]
    \label{eqn:v2}
  \end{equation}

  The ratio $R_\gamma$ is measured using the double ratio shown in Eqn. \ref{eqn:dubrat}.  It is defined as the ratio of the yield of inclusive photons to the yield of hadronic decay photons.  A ratio above one indicates a signal of direct photons.  The terms that go into the numerator of the double ratio, along with the correction factors, are shown in Eqns. \ref{eqn:pieceNum} and \ref{eqn:pieceDem}.  Note that all the terms are measured as a function of the converted photon $p_T$.  The numerator of Eqn. \ref{eqn:dubrat} is purely from data and consists of the ratio of the inclusive photon yield (the converted photons) to the yield of those photons that we tag as coming from $\pi^0$ decays.  It is necessary to measure the $\pi^0$ contribution since that is the largest source of hadronic decay photons.  This is done by reconstructing $\pi^0$'s by combining the converted photons with photons we measure in the EMCal.  To identify a photon in the EMCal we cut on the shower shape, comparing the shower shape of the photon candidate with the expected shower shape for a photon, and require a minimum energy deposition of 500 MeV.  The combinatorial background is removed with a mixed event technique and after this subtraction, we integrate the mass distrubution around $2\sigma$ of the $\pi^0$ peak.  In this way we (statistically) tag some of our inclusive photons as coming from a $\pi^0$ decay.  
  
  \begin{equation}
    R_\gamma = \frac{\gamma^{incl}(p_T)}{\gamma^{hadr}} = \frac{\epsilon_\gamma(p_T)f(p_T)\cdot\left(\frac{N^{incl}_\gamma(p_T)}{N^{\pi^0tag}(p_T)}\right)_{Data}}{\left(\frac{N^{hadr}_\gamma(p_T)}{N^{\pi^0}(p_T)}\right)_{Sim}}
    \label{eqn:dubrat}
  \end{equation}
  
  \begin{eqnarray}
    N^{incl}_\gamma(p_T) = c\epsilon_{pair}a_{pair}\gamma^{incl}(p_T)
    \label{eqn:pieceNum}      \\
    N^{\pi^0tag}_\gamma(p_T) = c\epsilon_{pair}a_{pair}\epsilon_\gamma f\gamma^{\pi^0}(p_T)
    \label{eqn:pieceDem}
  \end{eqnarray}
  
  In the Eqns. \ref{eqn:dubrat}, \ref{eqn:pieceNum}, and \ref{eqn:pieceDem}, $\gamma$ represents the yield of photons that nature produces, N represents the value we actually measure.  The correction factors for the inclusive photon yield are shown in Eqn. \ref{eqn:pieceNum}.  $\epsilon_{pair}$ and $a_{pair}$ are the $e^+e^-$ pair efficiency and acceptance in the PHENIX detector respectively.  There is also some factor corresponding to the probability of the photon externally converting, denoted by c.  These corrections apply to both the inclusive and the $\pi^0$ tagged photon samples.  For the $\pi^0$ tagged sample we additionaly have to correct for the reconstruction efficiency, $\epsilon_\gamma$, and apply an acceptance correction, f, for reconstructing the unconverted decay photon in the EMCal.  The f factor is a conditional acceptance correction and is defined as the probability of getting the unconverted photon in the EMCal, given that we already have the $e^+e^-$ pair from the other photon.  The $\epsilon_\gamma$ accounts for occupancy effects in the detector by embedding simulated $\pi^0$ decays into real events.  These corrections are in progress and are being studied with full GEANT3 based Monte Carlo simulations.
  We account for sources of decay photons other than the $\pi^0$ in the denominator of Eqn. \ref{eqn:dubrat} determined from simulations.  Main sources of decay photons are decays of mesons such as $\pi^{0}$, $\eta$, and $\eta'$, and are included in a photon cocktail.  The photon distributions from these decays are calculated in a decay generator, which realistically handles the kinematics of the decay.  The measured $\pi^0$ $p_T$ distribution from published data is fit with a Hagedorn function as is used as input to the photon cocktail for the $\pi^0$ $p_T$ shape.  The shape of the $p_T$ distributions for the other hadrons is determined by $m_T$ scaling, where the $p_T$ is replaced by $m_T = \sqrt{m + p_T}$, with m the mass of the hadron.  The relative hadron/$\pi^0$ ratios are also taken from actual measurement and input into the cocktail.  Then the denominator of the double ratio can be calculated, which is the ratio of the yield of all hadronic decay photons to the yield of decay photons coming from $\pi^0$s.
  
  The $\pi^0$ tagging technique will also be extended, allowing us to calculate the $\pi^0$ $v_2$ in a manner consistent with the rest of the analysis.

  \section{Results and Conclusions}
\begin{figure}[h]
    \begin{minipage}{18pc}
      \includegraphics[scale = 0.3]{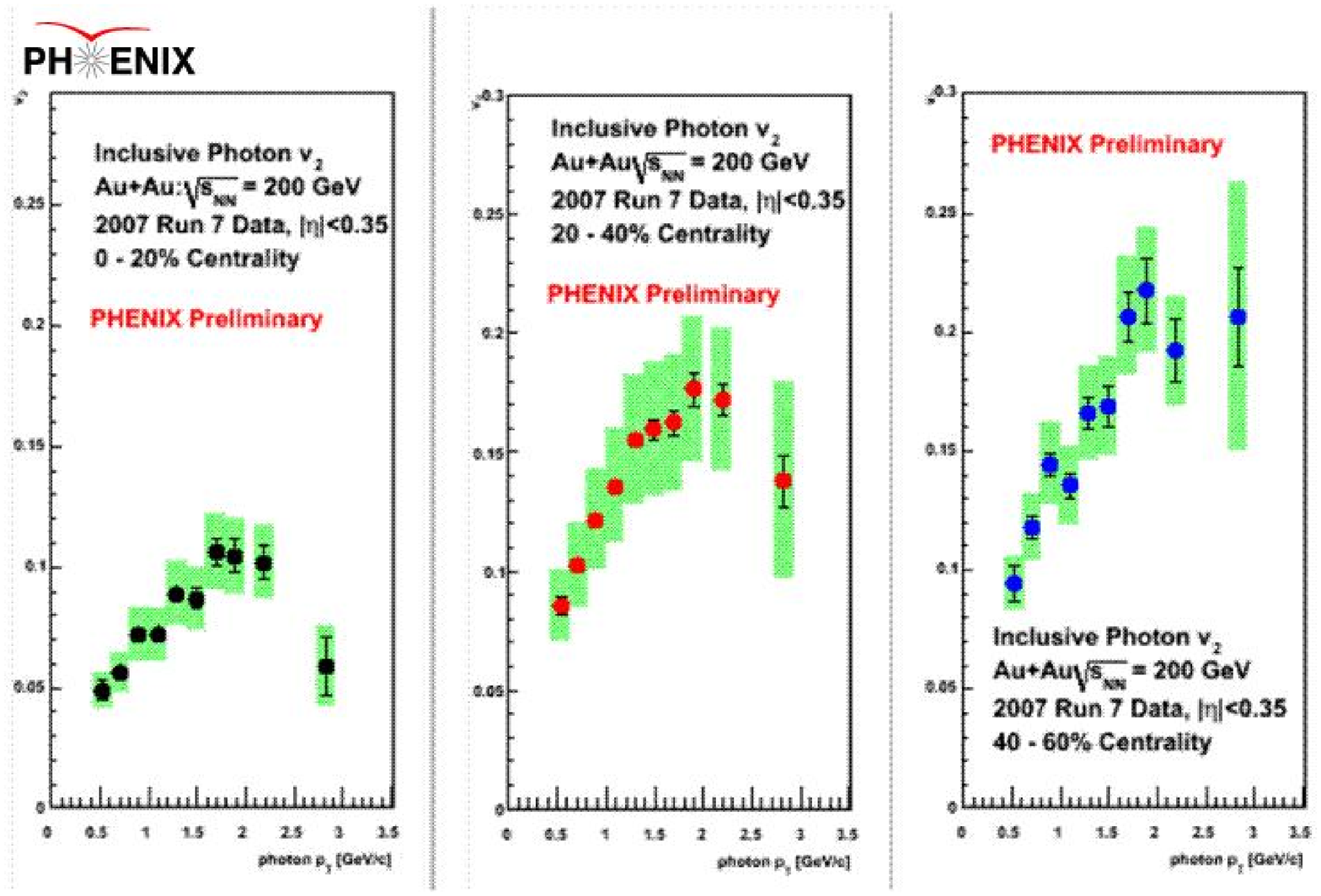}
      \caption{\label{fig:v2_prelim}The preliminary measured inclusive photon $v_2$ via the external conversion method.  This is shown in three centrality bins.  From left to right the bins are $0-20\%$, $20-40\%$, and $40-60\%$ centrality.}
    \end{minipage}\hspace{2pc}%
    \begin{minipage}{18pc}
      \includegraphics[scale = 0.38]{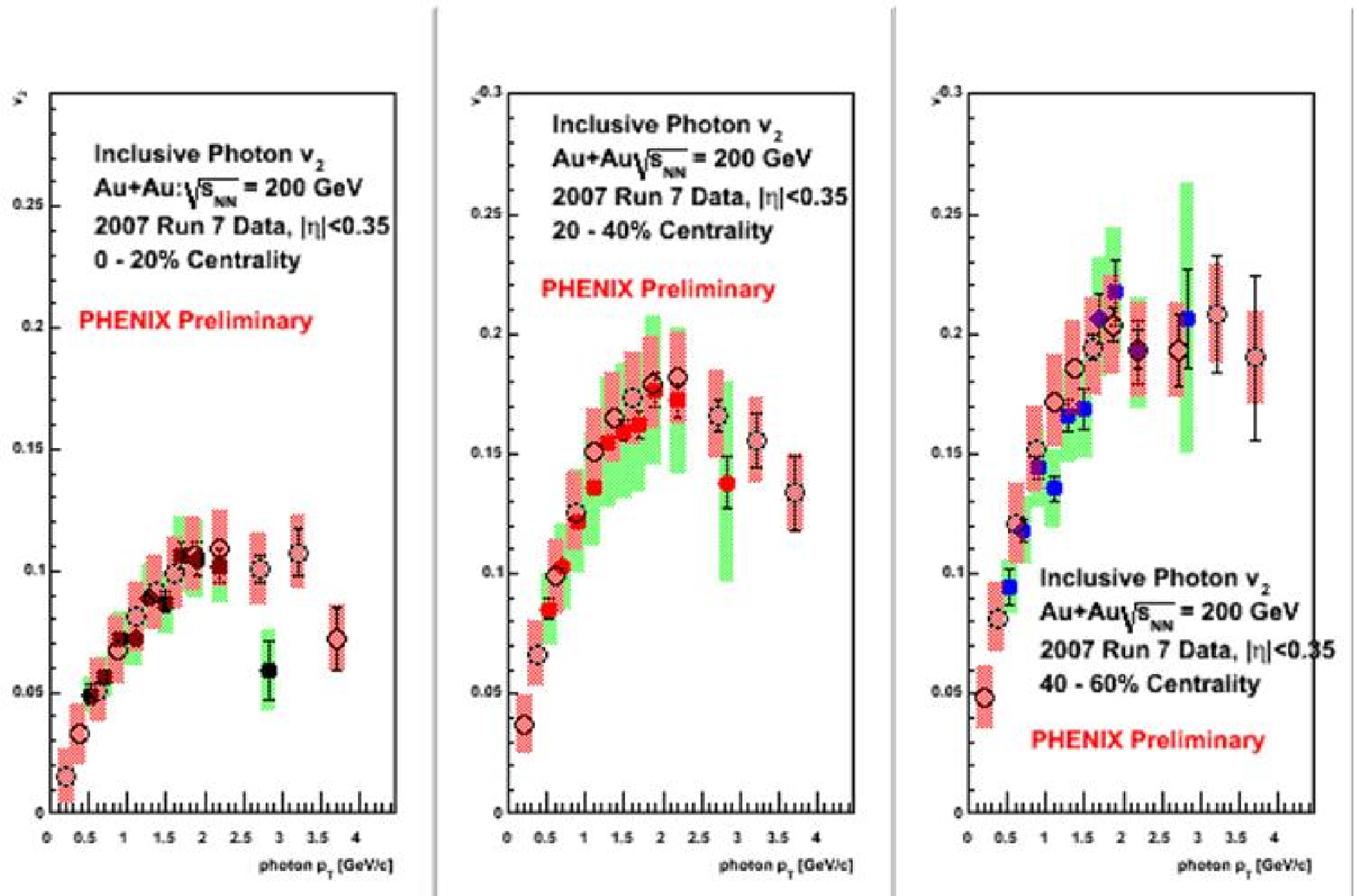}
      \caption{\label{fig:v2_compare}The same points as in \ref{fig:v2_prelim}, but with published data overlayed in the open black cirles with the red systematic error boxes~\cite{pub_v2}}
    \end{minipage} 
  \end{figure}

  This analysis is still a work in progress at the time of writing, therefore only the methodology and current status have been discussed.  A preliminary result has been obtained for the inclusive photon $v_2$, shown in Fig \ref{fig:v2_prelim}, for three centrality bins in a range $0.4 < p_T < 3.5$ GeV.  From left to right in the figure are the $0-20\%$, $20-40\%$, and $40-60\%$ centrality bins.  The major source of systematic error on the points comes from the reaction plane determination (about $10\%$), with photon id and calculation method contributing only a few percent.  The systematic error associated with the reaction plane resolution is common to both the inclusive photon $v_2$ and the $\pi^0$ $v_2$, thus when the final systematic error is propagated to the full direct photon $v_2$ result, the reaction plane systematic will only come in once.  These results are compared to published data~\cite{pub_v2} in which photons are directly measured in the EMCal.  The agreement is good, indicating the validity of this method.
  
  In summary, we have presented the current status and the analysis method for measuring the elliptic flow of direct photons via their external conversion to dilepton pairs in the detector material. A comparison of the measured inclusive photon $v_2$ with an earlier PHENIX result obtained using the EMCal shows the reliability of this method. In a next step, we will subtract the $v_2$ of decay photons and calculate the direct photon $v_2$, using the external conversion method.

\end{document}